\documentstyle[12pt]{article}
\begin{document}
\newcommand{\be}{\begin{equation}}
\newcommand{\ee}{\end{equation}}
\newcommand{\bea}{\begin{eqnarray}}
\newcommand{\eea}{\end{eqnarray}}
\def\pd{\partial}
\def\no{\nonumber}
%----------------------------------------- this is double spacing
\baselineskip=24pt plus 2pt
%----------------------------------------- this is double spacing
%%%%%%%%%%%%%%%%%%%%%%%%%%%%%%%%%%%%%%%%%%
\vspace{2mm}
\begin{center}
{\large \bf Stochastic Responses of the Stable Period-$p$ Orbits 
in One-Dimensional Noisy Map Systems } \\
\vspace{3mm}

Tian-Nan Wang\footnote{E-mail address: 
skyboy@ibm65.phys.ncku.edu.tw},
Rue-Ron Hsu\footnote{E-mail address: rrhsu@mail.ncku.edu.tw},
Han-Tzong Su\footnote{E-mail address: htsu@mail.ncku.edu.tw},
Wei-Fu Lin\footnote{E-mail address: 
wfl@ibm65.phys.ncku.edu.tw},\\
Jyh-Long Chern\footnote{E-mail address: 
jlchern@mail.ncku.edu.tw}, 
and Chia-Chu Chen\footnote{E-mail address: 
chiachu@mail.ncku.edu.tw}
\\
Nonlinear Science Group, Department of Physics,\\
National Cheng Kung University,\\ 
Tainan, Taiwan 70101, Republic of China\\
\end{center}
\vspace{2mm}
\begin{center}
{\bf ABSTRACT}
\end{center}  
~~~~We analytically derive the correlation functions of the 
stochastic response of period-$p$ orbits in a generic 
one-dimensional map system under the influence of weak 
external, additive white noise. Two approaches, a recurrence 
scheme 
and a path integral method, are provided.  Beside the supercycle, 
we recognise that the fluctuations on each elements of the 
period-$p$ 
orbits are colored noise which have non-vanishing time 
correlation. The results also indicate that the stochastic responses 
will be stationary after a large number of period-$p$ iterations. 
The 
analytical results are confirmed 
by numerical simulations.\\
PACS number(s): 02.50.Fz, 05.40.+j      \hfill{Typeset using \LaTeX}
\newpage 
%%%%%%%%%%%%%---section 1 ---%%%%%%%%%%%%%%%

\section{Introduction}

~~~~The role of dynamical noise 
in the deterministic nonlinear system has been studied 
extensively~\cite{1}-\cite{9}.  These results indicate that, in a 
noisy dynamical system, not only the higher periods become 
obliterated by the increasing noise level but the bifurcation points 
themselves become blurred.  In particular, at the points of 
bifurcation, the Lyapunov exponents no longer vanish as they do 
in the deterministic limit.  
Meanwhile, through both numerical and analytical studies, some 
universal scaling behaviors are discovered at the onset of chaos 
even in the noisy dynamical system~\cite{2}-\cite{5}.  Mayer-
Kress and Haken  carried out numerical investigation on the 
distribution function of stochastic response of a logistic model with 
uniform white noise using an integral recursion formula~\cite{6}. But 
the analytical form of the correlation function, or distribution 
function, of the 
stochastic response of period-$p$ orbits in a generic noisy map 
system was not  presented. 

Later on, 
in a review paper~\cite{7}, Crutchfield, 
Farmer and Huberman studied the stochastic response of a noisy 
one-dimensional map, $x_{n+1}=rf(x_n)+p_n$, where $f(x)=x(1-x)$ 
and $p_n$ is the external noise.  They proposed that the 
stochastic response should be regarded as the fluctuation on the 
control parameter $r$.  It means that the noisy map can be 
rewritten as $x_{n+1}=(r+q_n)f(x_n)$, where $q_n$ is the response 
parametric noise.  However, when we numerically study the 
bifurcation diagrams of the noisy logistic map and the noisy Ikeda 
map, 
see Fig. 1, we notice that the mean values of $x_n$ on each 
elements of noisy stable orbits coincide with those of the stable 
orbits without noise.  This fact implies that the dispersion should 
be regarded as the fluctuation on the elements of noise-free 
period-$p$ orbits $\{\bar x^{(p)}_m ~|~ 1\le m\le p \}$ of the 
one-dimensional map $x_{n+1}=f_r(x_n)$. 
In this map, the element of period-$p$ 
orbits $\bar x^{(p)}_m$ and control parameter $r$ obey the 
periodicity condition,
$ \bar x^{(p)}_m=f^p_r(\bar x^{(p)}_m),$
and the stability condition,
$|M^{(p)}|\equiv  |\prod^p_{m=1} (\frac {\partial f_r}{\partial 
x}\vert_{\bar x^{(p)}_m} )|<1.$

Based on this observation, 
we will derive the general formula for the 
correlation function of each elements on the period-$p$ orbits by 
a recurrence scheme and by a path integral method.  The results 
indicate that, 
beside the supercycles, the 
stochastic responses are colored in nature which means that the 
correlation 
between the $kp$-th step and the $(k+\ell)p$-th step are non-
vanishing. 
It also shows that the correlation functions, or the distribution 
functions, of the stochastic response are stationary after 
$kp$ iterations when $k>>1$.
The plan of this 
paper is as following. We derive the correlation functions by 
recurrence scheme in section 2, and by path integral method in 
section 3. Numerical verification on two interesting systems, 
logistic map and Ikeda map, are demonstrated in section 4.  
Finally, some concluding remarks are provided.

%%%%%%%%%%%%%---section 2 ---%%%%%%%%%%%%%%%

\section{Recurrence scheme}

~~~~Here, we consider a generic one-dimensional 
nonlinear 
map in which the external additive noise is involved dynamically, 
\be
x_{n+1}=f_r(x_n)+\eta_n,
\ee
where $\eta_n$ is white noise with zero mean, i.e. $<\eta_n 
\eta_m>=D\delta_{nm}$ and $<\eta_n>=0$. For examples, $D$ is 
$\frac 
{\sigma^2}{3}$ for uniform white noise with noise level $\sigma$ 
and $D=\sigma^2$ for Gaussian white noise with standard 
deviation $\sigma$.  

To derive the stochastic response on each elements of 
fluctuating period-$p$ orbit, we let the initial state $x_m$ to be 
located in the neighbourhood of
$\bar x^{(p)}_m$, i.e. $x_m=\bar x^{(p)}_m+\xi_m$, $\xi_m<<1$.  
After $kp$ iterations, the stochastic response on the $m$-th 
element 
of period-$p$ can be regarded as the fluctuation on the $m$-th 
element 
of noise-free period-$p$ orbits, 
that is 
\be
x_{kp+m}\equiv\bar x^{(p)}_{kp+m}+\xi_{kp+m}=\bar 
x^{(p)}_m+\xi_{kp+m}.
\ee 
Here, $\xi_{kp+m}$ is the stochastic response on the 
$m$-th elements of period-$p$ after $k$ times of period-$p$ 
iteration.

From now on, the noise level and stochastic response will be 
assumed to 
be small in comparison with the noise-free period-$p$ orbits such 
that the perturbation scheme is applicable.  By substituting Eq.(2) 
and $ x_{(k+1)p+m}=\bar x^{(p)}_m+\xi_{(k+1)p+m}$ into the 
$p$ times iteration of Eq.(1), i.e. 
\be
x_{(k+1)p+m}=f_r(...f_r(f_r(x_{kp+m})+\eta_{kp+m})+\eta_{kp+
m+1}...)+ \eta_{kp+m+p-1}. 
\ee
It leads to a recurrence relation between $\xi_{(k+1)p+m}$ 
and $\xi_{kp+m}$,
\be
\xi_{(k+1)p+m}=M^{(p)}\xi_{kp+m}+H^{(p-1)}_{k,m}.
\ee
Here, we have denoted 
\be
M^{(p)}\equiv\prod^{p-1}_{i=0}M^{(p)}_{m+i}
\ee 
and 
\be 
H^{(p-1)}_{k,m}\equiv \left \{ \begin{array}{ll}
\eta_{k+m} & \mbox{, for $p=1$,}\\
\\
\sum^{p-1}_{j=1}(M^{(p-1)}_{j,m}\eta_{kp+m+j-1})+ 
\eta_{kp+m+p-1} & \mbox{, for $p\ge 2$,}
\end{array} \right.  
\ee \\
where 
\be
M^{(p)}_{m+i}\equiv \frac {\partial f_r}{\partial x} \vert_{\bar 
x^{(p)}_{m+i}} 
\ee 
and 
\be
M^{(p-1)}_{j,m}\equiv \prod^{p-1}_{i=j}M^{(p)}_{m+i}.
\ee
By iterating the recurrence relation, Eq.(4), one obtains 
\be
\xi_{kp+m}=(M^{(p)})^k\xi_m+\sum_{i=1}^k (M^{(p)})^{i-1}H^{(p-
1)}_{k-i,m}.
\ee

For simplicity, but without loss of generality, we set the initial 
state to be
$x_m=\bar x^{(p)}_m$, that is $\xi_m=0$. Therefore, the stochastic 
response, after 
$kp$ times of iteration, becomes
\be
\xi_{kp+m}=\sum_{i=1}^k (M^{(p)})^{i-1}H^{(p-1)}_{k-i,m}.
\ee
As a result, the mean fluctuation of stochastic response is zero 
when the $<\eta_k>=0$. Here, $<\eta_k>\equiv 
{\lim_{N\rightarrow\infty}}\frac{1}{N}\sum^{N-1}_{i=0} 
\eta^{(i)}_{k}$ denotes the ensemble average. The correlation 
function of stochastic response between the $kp$-th step and the 
$(k+\ell)p$-th
step can be written as
\be
<\xi_{(k+\ell)p+m}\xi_{kp+m}>=\sum_{i'=1}^{k+\ell}\sum_{i=1}^k  
(M^{(p)})^{i'-1}(M^{(p)})^{i-1}<H^{(p-1)}_{(k+\ell)-i',m}
H^{(p-1)}_{k-i,m} >.
\ee 
Using the properties of the input white noise, we find, 
for the case $p\ge 2$,
\bea
&&< H^{(p-1)}_{(k+\ell)-i',m}H^{(p-1)}_{k-i,m}>\no\\
&&=
\sum^{p-1}_{j'=1}\sum^{p-1}_{j=1} M^{(p-1)}_{j',m}M^{(p-1)}_{j,m} 
<\eta_{(k+\ell-i')p+m+j'-1}\eta_{(k-i)p+m+j-1}> \no\\
&& +<\eta_{(k+\ell-i')p+m+p-1}\eta_{(k-i)p+m+p-1}> \no\\
&&=
\sum^{p-1}_{j'=1}\sum^{p-1}_{j=1} M^{(p-1)}_{j',m}M^{(p-1)}_{j,m} 
\delta_{j,j'}\delta_{i,i'-\ell}D+ \delta_{i,i'-\ell}D\no\\
&&=
\Bigl(1+\sum_{j=1}^{p-1}(M^{(p-1)}_{j,m})^2\Bigr) \delta_{i,i'-
\ell}D.
\eea
Therefore, the correlation function ( for $p\ge 2$ ) between the $kp$-th step and 
the $(k+\ell)p$-th step is
\be
<\xi_{(k+\ell)p+m}\xi_{kp+m}>=\frac {(M^{(p)})^\ell (1-
(M^{(p)})^{2k})}
{1-(M^{(p)})^2} \Bigl\{1+\sum_{j=1}^{p-1}(M^{(p-1)}_{j,m})^2 
\Bigr\} D.
\ee 
It should be noted that the term 
$\sum_{j=1}^{p-1}(M^{(p-1)}_{j,m})^2$ in Eq.(13) will be zero for the most trivial case $p=1$.
%%%%%%%%%%%%%---section 3 ---%%%%%%%%%%%%%%%

\section{Path integral approach}

~~~~As it has been shown in previous section 
using the recurrence scheme,  one can derive
 the correlation function of stochastic response of a dynamical 
map contaminated by uniform or Gaussian 
white noise.  Due to the excellent form of Gaussian 
white noise
\be
\rho (\eta_i)=\frac {1} {\sqrt {2\pi\sigma^2}}exp\Bigl\{-\frac 
{\eta_i^2}{2\sigma^2}\Bigr\},
\ee
one also can derive its stochastic response by using  
path integral approach.  

For simplicity, we assume the initial state is located at 
$x_m=\bar x_m^{(p)}$, it means that the probability distribution 
of $x$ is ${\it P}(x_m)=\delta(x_m-\bar x_m^{(p)})$ at the 
beginning.
After one iteration, $x_{m+1}=f_r(x_m)+\eta_m$, the distribution 
function of $x_{m+1}$ disturbed by the Gaussian noise has become
\bea
{\it P}(x_{m+1})&=&\int dx_m \rho (x_{m+1}-f_r(x_m)){\it
P}(x_m)\\
&=&\frac {1} {\sqrt {2\pi\sigma^2}}exp\Bigl\{-\frac {(x_{m+1}-
f_r(\bar x_m^{(p)}))^2}{2\sigma^2}\Bigr\}. \no
\eea
After the next iteration, 
$x_{m+2}=f_r(x_{m+1})+\eta_{m+1}$, one can write down the 
distribution 
function of $x_{m+2}$ 
\be
{\it P}(x_{m+2})=\int dx_m \int dx_{m+1} \rho (x_{m+1}-
f_r(x_m))\rho (x_{m+2}-f_r(x_{m+1})){\it P}(x_m).
\ee
Now, we  introduce the kernel $K(x_{m+\ell},x_m)$ which is 
defined as
\be
{\it P}(x_{m+\ell})=\int dx_m K(x_{m+\ell},x_m){\it P}(x_m).
\ee
The kernel can be read out from Eq.(16) and (17), that is 
\be
K(x_{m+\ell},x_m)=\int \prod_{j=1}^{\ell-1} dx_{m+j} 
\prod_{j=0}^{\ell-1} K(x_{m+j+1},x_{m+j}), 
\ee
where 
\be
K(x_{m+j+1},x_{m+j})=\frac {1} {\sqrt {2\pi\sigma^2}}exp\Bigl\{-
\frac {(x_{m+j+1}-f_r(x_{m+j}))^2}{2\sigma^2}\Bigr\}.
\ee
Then, the distribution function of $x_{kp+m}$, after $kp$ 
iterations 
begins with $x_m=\bar x_m^{(p)}$, is
\be
{\it P}(x_{kp+m})=\int \prod^{kp-1}_{j=0} \Bigl\{ dx_{m+j} \frac 
{1} {\sqrt {2\pi\sigma^2}}exp\bigl\{-\frac {(x_{m+j+1}-
f_r(x_{m+j}))^2}{2\sigma^2}\bigr\}\Bigr\}\delta (x_m-\bar 
x_m^{(p)}).
\ee

To obtain the distribution function by integration, we change the 
variable 
$x_{m+j}$ to $x_{m+j}=\bar x^{(p)}_{m+j}+\xi_{m+j}$.  As 
mentioned in section 2, $\xi_{m+j}$ is the corresponding stochastic 
response and is assumed to be small enough that the linear 
perturbation scheme is applicable.  Therefore, we have 
$dx_{m+j}=d\xi_{m+j}$ 
and 
\bea
f_r(x_{m+j}) &=& f_r(\bar x^{(p)}_{m+j})+\frac {\partial 
f_r}{\partial x}\Big\vert_{\bar 
x^{(p)}_{m+j}}\xi_{m+j}+O(\xi^2)\no\\
&\approx&\bar x^{(p)}_{m+j+1}+M^{(p)}_{m+j} \xi_{m+j}, 
\eea
and the distribution function can be expressed as
\be
{\it P}(x_{kp+m})=\int \prod^{kp-1}_{j=0} \Bigl\{ d\xi_{m+j} \frac 
{1} {\sqrt {2\pi\sigma^2}}exp\bigl\{-\frac {(\xi_{m+j+1}-
M^{(p)}_{m+j} \xi_{m+j})^2}{2\sigma^2}\bigr\}\Bigr\}\delta 
(\xi_m).
\ee
Using the formula 
\bea
\int &dz &\frac {1} {\sqrt {2\pi\sigma_1^2}} exp\Bigl\{-\frac {(x-
az)^2}{2\sigma_1^2}\Bigr\} \frac {1} 
{\sqrt {2\pi\sigma_2^2}} exp\Bigl\{-\frac {(z-
by)^2}{2\sigma_2^2}\Bigr\}\no\\ 
&=& \frac {1} {\sqrt {2\pi(\sigma_1^2+a^2\sigma_2^2)}} 
exp\Bigl\{-\frac {(x-aby)^2}{2(\sigma_1^2+a^2\sigma_2^2)}\Bigr\},
\eea
and integrating out $d\xi_m ...d\xi_{(kp-1)+m}$ step by step.
Finally, we have 
\be
{\it P}(x_{kp+m})=\frac {1} {\sqrt {2\pi\sigma_{kp+m}^2}} 
exp\Bigl\{-\frac {(x_{kp+m}-\bar x^{(p)}_m )^2} 
{2\sigma_{kp+m}^2}\Bigr\},
\ee
where 
\be
\sigma_{kp+m}^2=\sigma^2\Bigl( \frac {1-(M^{(p)})^{2k}} {1-
(M^{(p)})^{2}} \Bigr) \Bigl\{1+\sum_{j=1}^{p-1}(M^{(p-1)}_{j,m})^2 
\Bigr\}.
\ee
This result indicates that the stochastic response on each elements 
of period-$p$ orbit  for the input Gaussian noise still is a 
Gaussian-like distribution with mean $\bar x^{(p)}_m$ and 
deviation $\sigma_{kp+m}$. This is consistent with the case 
$\ell=0$ of Eq.(13). 

It is expected that the stochastic response will inherit the time 
correlation from the deterministic dynamical system.  Here, we 
define the time correlation between the $kp$-th step and the 
$(k+\ell)p$-th step 
of stochastic responses as
\bea
&&<(x_{(k+\ell)p+m}-<x_{(k+\ell)p+m}>)(x_{kp+m}-
<x_{kp+m}>)>\no\\
&&\equiv <\xi_{(k+\ell)p+m}\xi_{kp+m}>\no\\
&&=\int d x_{(k+\ell)p+m}\int d x_{kp+m} 
\xi_{(k+\ell)p+m}\xi_{kp+m} K(x_{(k+\ell)p+m},x_{kp+m}) 
{\it P}(x_{kp+m}).\no\\
\eea
From Eq.(18) and Eq.(23), one finds
\be
 K(x_{(k+\ell)p+m},x_{kp+m})=\frac {1} {\sqrt {2\pi\sigma_{\ell 
p}^2}} exp\Bigl\{-\frac {(\xi_{(k+\ell)p+m}-
(M^{(p)})^{\ell}\xi_{kp+m})^2} {2\sigma_{\ell p}^2}\Bigr\},
\ee
where 
\be
\sigma_{\ell p}^2=\sigma^2\Bigl( \frac {1-(M^{(p)})^{2\ell}} {1-
(M^{(p)})^{2}} \Bigr) \Bigl\{1+\sum_{j=1}^{p-1}(M^{(p-1)}_{j,m})^2 
\Bigr\}.\no
\ee
Therefore, the correlation between the $(k+\ell)p$-th step and 
the $kp$-th step is 
\be
<\xi_{(k+\ell)p+m}\xi_{kp+m}>=(M^{(p)})^{\ell}\sigma_{kp+m}^{2}.
\ee
These results are exactly the same as Eq.(13) which is obtained 
from the
recurrence scheme.

%%%%%%%%%%%%%---section 4 ---%%%%%%%%%%%%%%%

\section{Numerical verification}

~~~~To check the correctness of our analytical derivation, 
we compare the analytical results to the numerical results in two 
interesting systems: logistic map and Ikeda map.  
Firstly, let us consider the noisy logistic map which is 
contaminated by uniform white noise with amplitude 
$\sigma=0.001$.  Fig.2(a) shows that the ratio between the 
standard deviation of stochastic response and input noise, 
$\sqrt{\frac {<\xi_{kp+m}\xi_{kp+m}>}{D}}$, on each $m$-th 
element 
of period-1, 2, 4 orbits in the noisy logistic map, $x_{n+1}=rx_n(1-
x_n)+\eta_n$,  for $1.0\le r\le 3.54$.  The normalized correlation 
function between the $(k+\ell)p$-th step and the $kp$-th step , 
$\sqrt{\frac 
{<\xi_{(k+\ell)p+m}\xi_{kp+m}>}{D}}$, on each $m$-th element of 
period-2 orbit in the logistic map for $\ell=1,2$, are presented in 
Fig. 2(b) and 2(c), respectively.  Similar comparison for 
the case of the noisy Ikeda map, $x_{n+1}=1-r(1-cos(\frac 
{\pi}{2}x_n))+\eta_n$,  are 
demonstrated in Fig.3(a)-(c).

When the Gaussian white noise is added into the dynamical 
system, it is worth examining the distribution function ${\it
P}(x_{kp+m})$ directly.  Fig.4(a)-(c) show the numerical and 
analytical distribution functions ${\it P}(x_{kp+m})$ of $m$-th 
element of period-2 orbit in the logistic model with control 
parameters 
$r=3.1$, $1+{\sqrt 5}$ and $3.3$, respectively.

As one can see from the graphs, all the numerical results agree 
with the analytical derivation very 
well.  So long as the ratio between the amplitude of the input 
noise 
and 
the value of fixed point is less than $5\%$, and the dynamical 
system is not too close to the bifurcation point.

%%%%%%%%%%%%%---section 5 ---%%%%%%%%%%%%%%%

\section{Concluding remarks}

~~~~From the derivation of the general formula for the 
correlation function of each elements on the period-$p$ orbits by 
the recurrence scheme and the path integral method, 
we know that the 
recurrence approach is good for any white noise with zero mean, 
but path integral approach is only feasible for Gaussian white 
noise.  
However, the path integral method give us the distribution 
function, with this, one can easily to write down the higher 
moments 
of correlation function directly.   Beside the supercycles 
$(M^{(p)}=0)$, the stochastic response on each elements of 
period-$p$ orbit are colored which have the correlation time 
$\tau^{(p)}_c=-\frac {1} {\ln|M^{(p)}|}$. This time correlation can be 
regarded as an inheritance from the deterministic properties of 
the dynamical system.  

The general formula Eq.(13) also indicates that the amplitude of 
stochastic response is always larger than the level of input noise 
. The response on the supercycle is the most attractive one that 
the 
response of each element is independent of $k$, the times of 
period-$p$ iteration, that is
\be
<\xi_{kp+m}\xi_{kp+m}>=\Bigl\{1+\sum_{j=1}^{p-1}(M^{(p-
1)}_{j,m})^2 \Bigr\} D.
\ee 
The minimum response occurs at the $m=1$ elements on the 
supercycles in which $M^{(p-1)}_{j,1}=0$ and the response is 
identical to the original input white noise, see Fig.2(a) and Fig.3(a).  

Moreover, from the stability condition $|M^{(p)}|<1$,  one can 
deduce that 
\be
<\xi_{kp+m}\xi_{(k+\ell)p+m}> \approx \frac {(M^{(p)})^\ell }
{1-(M^{(p)})^2} \Bigl\{1+\sum_{j=1}^{p-1}(M^{(p-1)}_{j,m})^2 
\Bigr\} D,
\ee 
for $k>>1$. It means that after a large number of period-$p$ 
iterations 
the correlation functions on each elements become stationary.  
Therefore, when one defines the long-time average after large $k$ 
(transient) as 
\be
<\xi_{kp+m}\xi_{kp+m}>_{LT}\equiv \lim_{N \rightarrow \infty} 
\frac{1} {N} \sum^{N-1}_{i=0} \xi_{(k+i)p+m}\xi_{(k+i)p+m}.
\ee
The correlation function almost matches that computed by the 
ensemble average. 
Fig.5 
shows that Eq.(30) is a good approximation for the correlation 
function evaluated by the long-time average of the stochastic 
response of logistic map after the transient steps $k=1,000$ 
, if the neighborhood of the bifurcation point is excluded. This 
approximation 
has been reported in our previous papers~\cite{10}~\cite{11}. 
Indeed, those numerical results also indicate that the noise leads 
to a shift of the bifurcation points. This feature could account for 
the failure of our 
schemes near the bifurcation points.  It may imply that one 
should regard the response of noise as parametric noise in the 
neighborhoods of the bifurcation points, such as Crutchfield had 
done before, or deal with the response by other method~\cite{12}.

As an application, we have used the recurrence scheme to 
study the stochastic response of colered noise and the noise 
cascading in uni-directionally coupled elements system~\cite{11}. 
Those results offer us an opportunity to develop suitable methods 
to control noise in the system, see reference [11].  

The recurrence method can be extended to study the 
stochastic response of periodic orbit in continuous time system, as 
long as one can write down the Poincar\'e map of the dynamical 
system. In fact, Svensmark and Samuelsen~\cite{12}  had 
obtained a similar result for period-1 case in a noisy dynamical 
system, see Eq.(11) in reference [12].  They also found other 
type of auto-correlation function in a noisy dynamical 
system under the influence of a period-2 resonant perturbation.
On the other hand, using the Floquet theory, Wiesenfeld~\cite{13} 
obtained the noise response of a special model of continuous time 
system as an integral of noise source and fundamental matrix, and 
give a generic correlation function for the period-1 orbit.  Both papers focus on the influence of noise in the vicinity of 
period-doubling, but pay little attention on the correlation 
function for each elements of period-$p$ orbits.

{\sl Note in addition:}
After the completion of this manuscript, Prof. Weiss informed us his related works published about a decade ago~\cite{14}.  In his paper, he calculated the first and the second moments of the probability distribution for a N-dimensional noisy map, up to the second order approximation. The  $\ell=0$ case in our results is a special case, the first order approximation of one-dimensional noisy map, of his result in slightly different form, but he did not state the response of each elements on the stable orbits explicitly.  Moreover, he did not compute the time correlation function between the $kp$-th step and $(k+\ell)p$-th step for the case $\ell\neq 0$, that reveals the inheritance of the dynamical property from the deterministic structure.

\vspace{20mm}

\begin{center}

{\large \bf Acknowledgements}

\end{center}

\vspace{10mm}
We thank professor J. B. Weiss who informed us his excellent work . This work is partially supported by the National Science Council,
Taiwan, R.O.C. under the contract numbers. 
NSC 87-2112-M006-010, 
NSC 87-2112-M006-011, and NSC 87-2112-M006-0012.

\newpage

\begin{flushleft}

{\Large\bf Figure Captions :}\\

\end{flushleft}
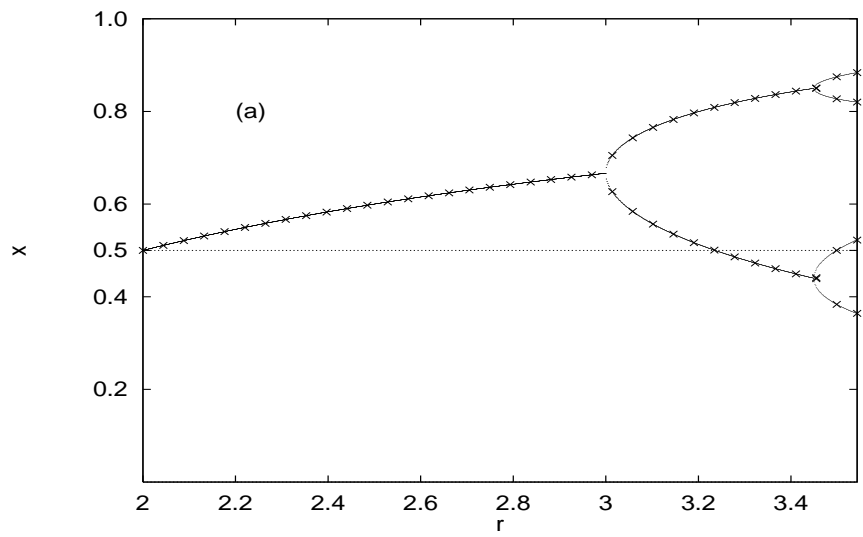
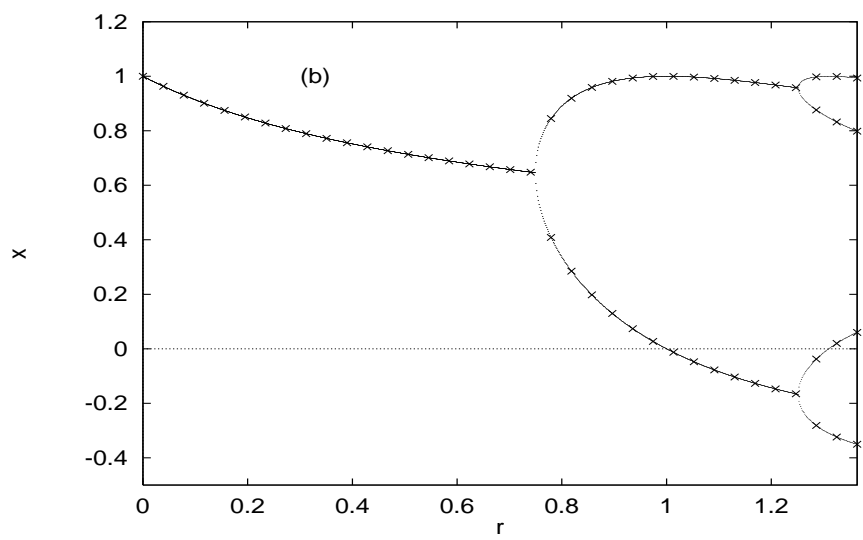
{\bf Figure 1.}

The coincidence between the mean value of each elements of 
noisy stable orbit and those stable orbits without noise, are 
demonstrated in (a) noisy 
logistic map, $x_{n+1}=rx_n(1-x_n)+\eta_n$, and (b) noisy Ikeda 
map, $x_{n+1}=1-r(1-cos(\frac {\pi}{2}x_n))+\eta_n$.  The 
amplitude of white noise $\eta_n$ is 0.001.  The cross ($\times$) 
denotes the average for 10,000 ensembles on each elements of 
period-$p$ orbit after 1,000 times of period-$p$ iteration. The 
solid line denotes the elements of noise-free orbit .
\\
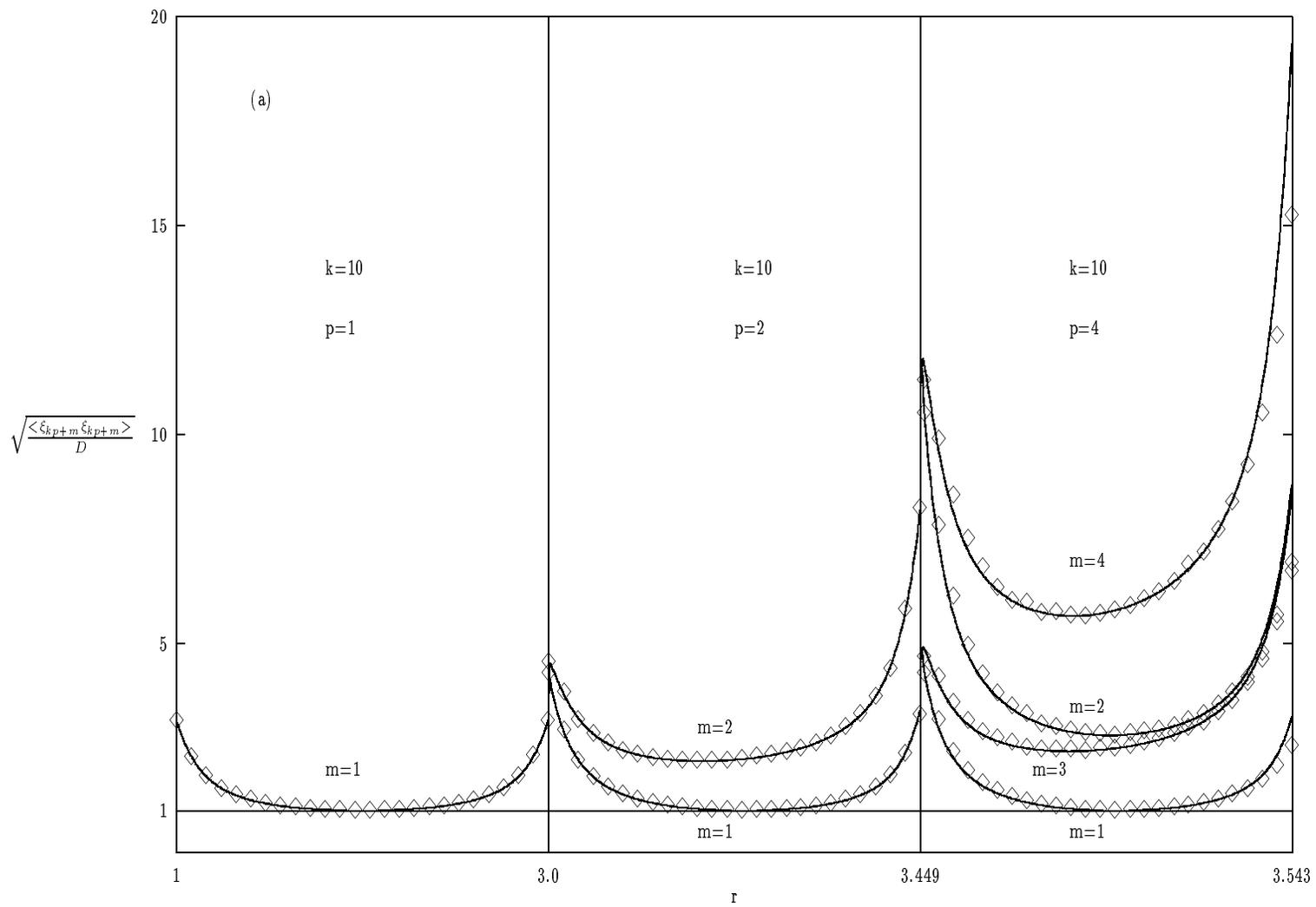
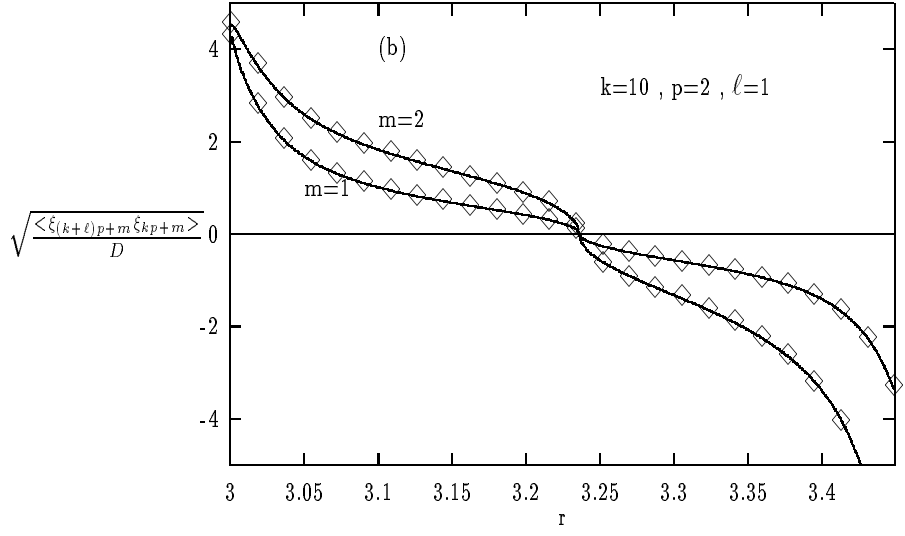
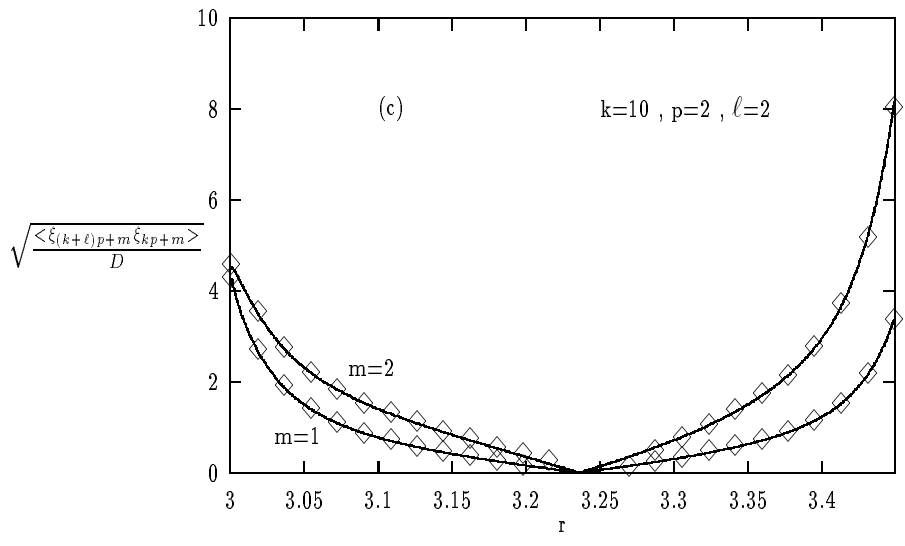
{\bf Figure 2.} 
 
$\sqrt{\frac {<\xi_{(k+\ell)p+m}\xi_{kp+m}>}{D}}$ via $r$ for 
$k=10$, 
$\ell=0,1,2$ are shown in (a), (b) and (c), respectively, see text.  
Here, 
the dynamical system is the noisy logistic map with noise 
amplitude $\sigma=0.001$. The diamond ($\diamond$) denotes 
the numerical simulation for 10,000 ensemble and the solid lines 
are the analytical results.
\\
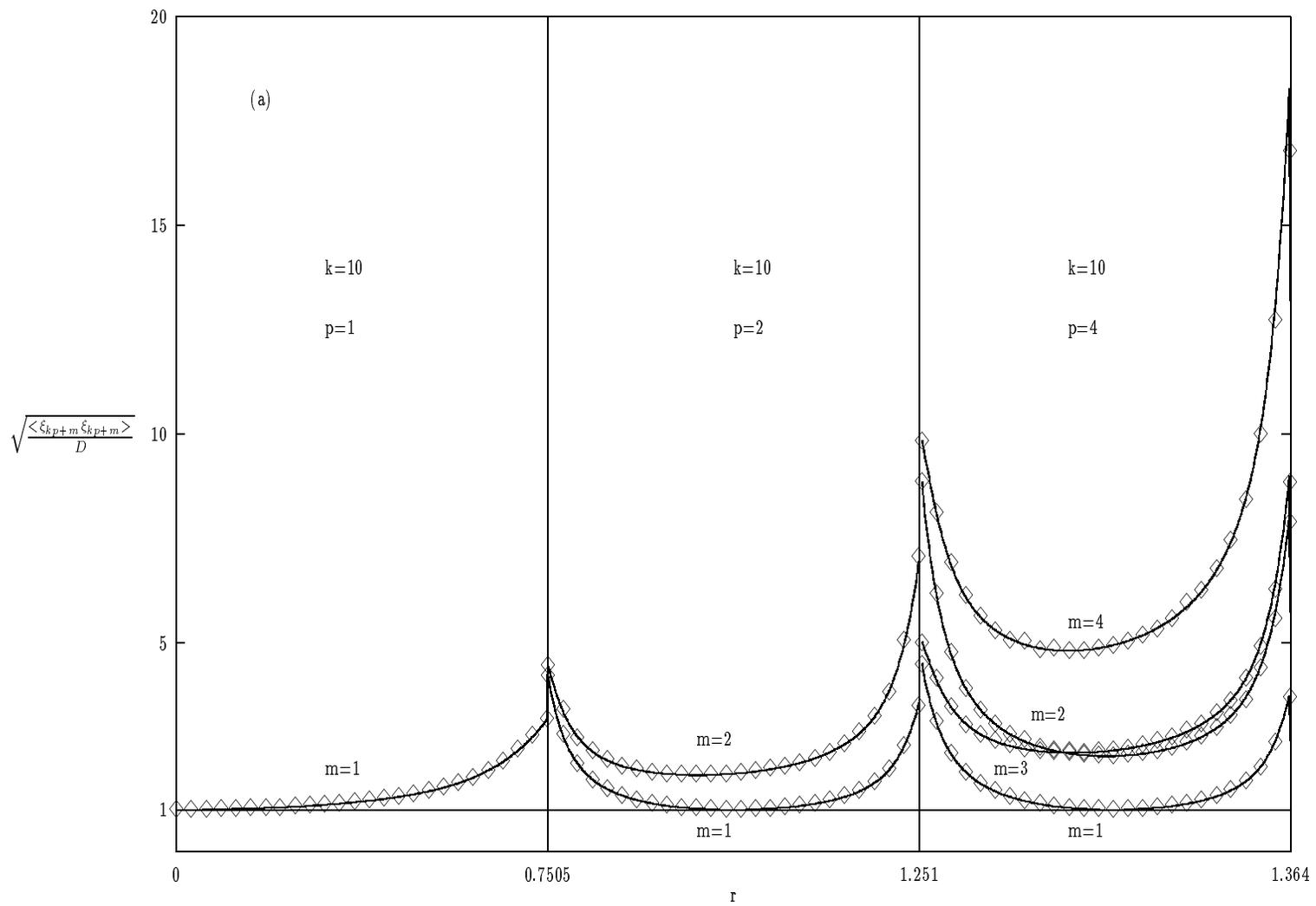
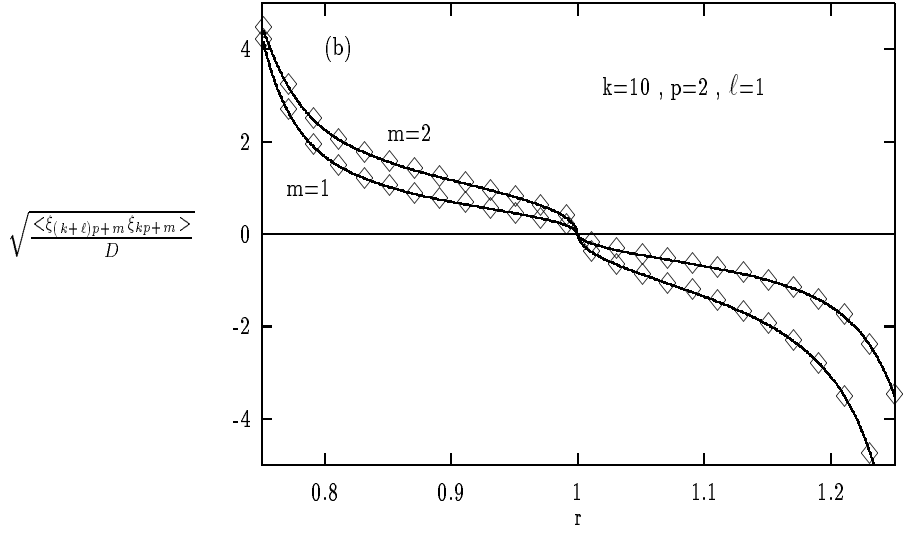
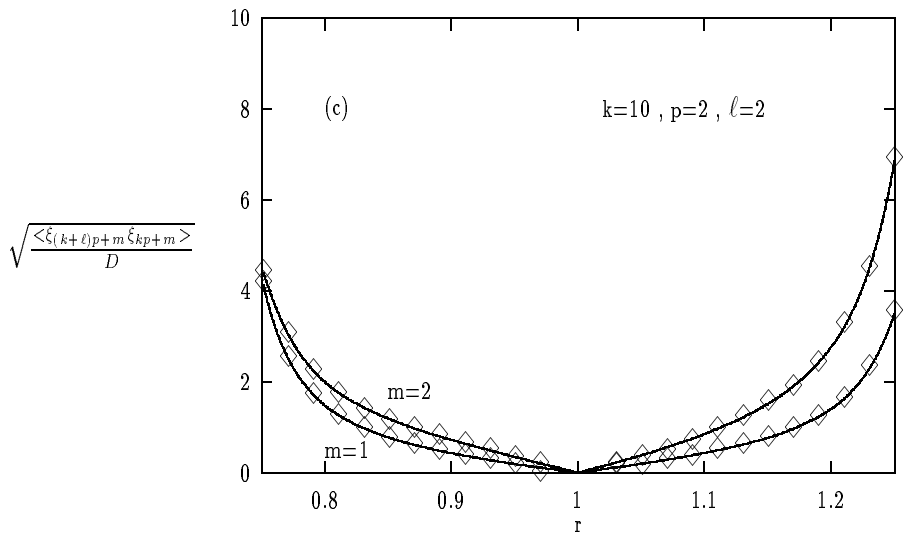
{\bf Figure 3.}  

Similar to Fig.2 (a)-(c), but the Ikeda map 
$x_{n+1}=1-r(1-cos(\frac {\pi}{2}x_n))+\eta_n$, go instead of the 
logistic map. 
\\
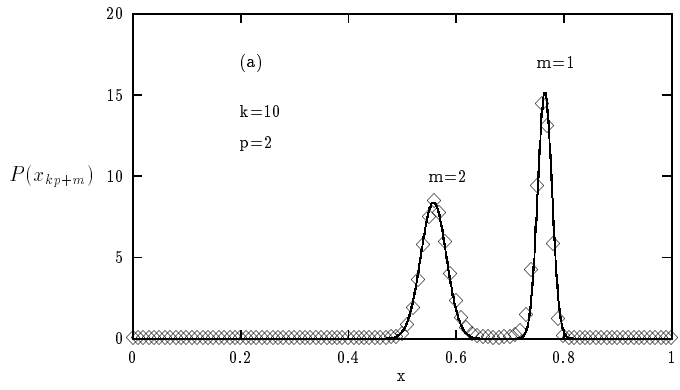
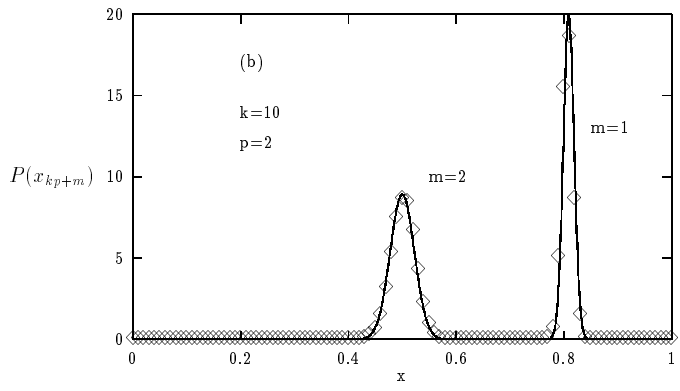
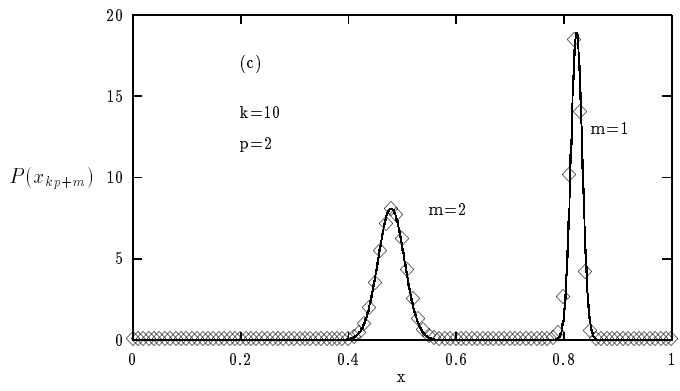
{\bf Figure 4.} 
 
The distribution function ${\it P}(x_{kp+m})$ of the $m$-th 
element of period-2 in logistic map with control parameters 
(a)$r=3.1$, (b)$1+{\sqrt 5}$ and (c) $r=3.3$.  Here, $k=10$ and the 
standard deviation of input Gaussian white noise is 
$\sigma=0.1$.  The diamond ($\diamond$) denotes the numerical 
simulation and the solid lines are analytical results. 
\\
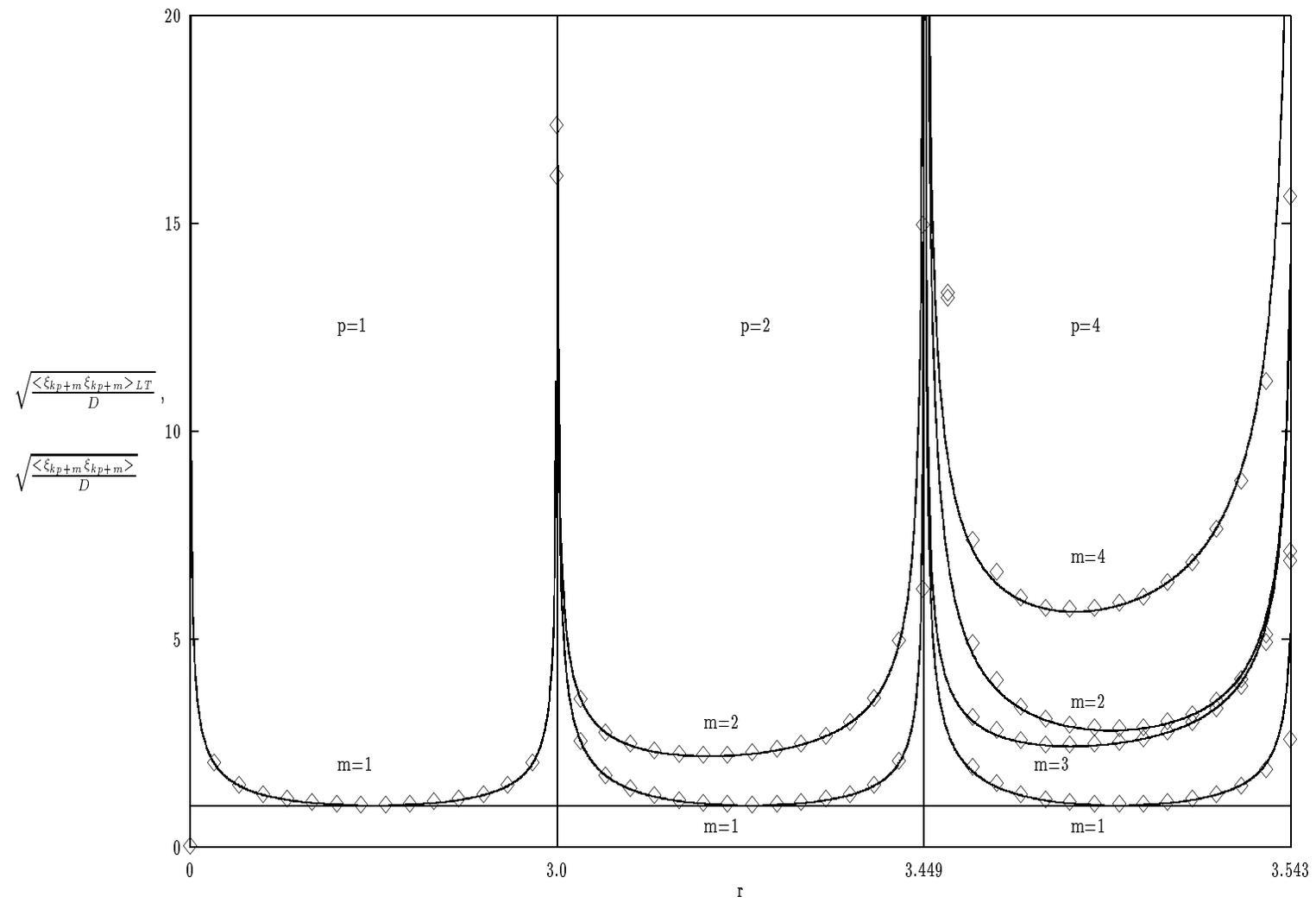
{\bf Figure 5} 

The diamond ($\diamond$) denotes the numerical results of 
correlation function evaluated by long-time average for different 
$r$ in logistic map.  Here, we take 50,000 data for the long-time 
average after 1,000 transient steps.  The solid lines are the 
analytical results of 
Eq.(30) for $k=1,000$, $\ell=0$.
\\


\newpage



\begin{thebibliography}{set}

\bibitem{1}
E. N. Lorenz, {\bf Ann. NY Acad. Sci. 357}, 282 (1980).
\bibitem{2}
J.P. Crutchfield and Huberman, {\bf Phys. Lett} {\bf A77}, 407 
(1980). 
\bibitem{3}
J.P. Crutchfield, M. Nauenberg and J. Rudnick {\bf Phys. Rev. Lett.}
{\bf 46}, 933 (1981). 
\bibitem{4}
B. Shraiman, C. E. Wayne and P. Martin, {\bf Phys. Rev. Lett.}
 {\bf 46}, 935 (1981). 
\bibitem{5} 
M.J. Feigenbaum and B. Hasslacher, {\bf 
Phys. Rev. Lett.} {\bf 49}, 605 (1982). 
\bibitem{6}
G. Mayer-Kress and H. Haken, {\bf J. Stat. Phys.}
 {\bf 26}, 149 (1981). 
\bibitem{7}
J.P. Crutchfield, J.D.Farmer and B.A. Huberman, 
{\bf Phys. Rep.} {\bf 92}, 45 (1982).
\bibitem{8}
H. G. Schuster, {\sl "Deterministic Chaos: an introduction "}, 3rd ed., 
(Weinheim; New York; Basel; Cambridge; Tokyo; VCH, 1995).
\bibitem{9}
T. Kapitaniak, {\sl "Chaos in systems with noise"}, 2nd ed., (World 
Scientific, Singapore,1990).
\bibitem{10}
R.-R. Hsu, H.-T. Su, J.-L. Chern, and C.-C. Chen, {\bf Phys. Rev. Lett.}
 {\bf 78}, 2936 (1997). 
\bibitem{11}
R.-R. Hsu, J.-L. Chern, W.-F. Lin, and C.-C. Chen, preprint: chao-
dyn/9710008.
\bibitem{12}
H. Svensmark and M. R. Samuelsen, {\bf Phys. Rev. A}
 {\bf 36}, 2413 (1987). 
\bibitem{13}
K. Wiesenfeld, {\bf J. Stat. Phys.}
 {\bf 38}, 1071 (1985). 
\bibitem{14}
J. B. Weiss, {\bf Phys. Rev. A}
 {\bf 35}, 879 (1987). 

\end{thebibliography}
\end{document}